\documentclass[12pt,a4paper]{article}

\pdfoutput=1

\usepackage[utf8x]{inputenc}
\usepackage[T1]{fontenc}
\usepackage{lmodern}
\usepackage[english]{babel}

\usepackage[rmargin=2.5cm,
            lmargin=2.5cm,
            top=2.5cm,
            bottom=2.5cm]{geometry}
            
\usepackage{enumerate}
\usepackage{paralist}
\renewcommand{\labelenumi}{(\theenumi)}

\usepackage{graphicx}
\usepackage{subfig}
\usepackage{placeins}

\usepackage{amsmath}
\usepackage{amsfonts}
\usepackage{amssymb}
\usepackage{dsfont}

\usepackage{longtable}
\usepackage{multirow}

\usepackage[colorlinks=false, pdfborder={0 0 0}]{hyperref}

\usepackage{natbib}
\usepackage{url}

\usepackage{caption}
\usepackage{framed}

\newcommand{\vectheta}{\boldsymbol{\theta}}
\newcommand{\vecphi}{\boldsymbol{\phi}}
\newcommand{\vecgamma}{\boldsymbol{\gamma}}

\newcommand{\by}{\boldsymbol{y}}
\newcommand{\bY}{\boldsymbol{Y}}

\newcommand{\Prob}{\mathbb{P}}
\newcommand{\E}{\mathbb{E}}

\newcommand{\pnorm}{N}
\newcommand{\pinvgamma}{IG}

\begin{document}

\begin{center}

{\Large \bfseries Bayesian Exponential Random Graph Models with Nodal Random Effects}
\vspace{5 mm}

{\large S. Thiemichen$^\star$, N. Friel$^\dagger$, A. Caimo$^\ddagger$, G. Kauermann$^\star$. } \\
{\textit{$^\star$Institut f\"ur Statistik  Ludwigs-Maximilians-Universit\"at M\"unchen, Germany.\\
$^\dagger$School of Mathematical Sciences and Insight: The National Centre for Data Analytics, \\ University College Dublin, Ireland.\\
$^\ddagger$Social Network Analysis Research Center, Faculty of Economics, University of Lugano, Switzerland.}}

\vspace{5 mm}

\today 

\vspace{5mm}

\end{center}

\begin{abstract}
We extend the well-known and widely used Exponential Random Graph Model (ERGM) by including nodal random effects to compensate for heterogeneity in the nodes of a network. The Bayesian framework for ERGMs proposed by \cite{CaimoFriel:2011} yields the basis of our modelling algorithm. A central question in network models is the question of model selection and following the Bayesian paradigm we focus on estimating Bayes factors. To do so we develop an approximate but feasible calculation of the Bayes factor which allows one to pursue model selection. 
Two data examples and a small simulation study illustrate our mixed model approach and the corresponding model selection.
\end{abstract}

\section{Introduction}
\label{sec:Introduction}

The analysis of network data is an emerging field in statistics which is challenging both model-wise and computationally. Recently \cite{GoldenbergZheng:2010}, \cite{HunterKrivitsky:2012}, \cite{Fienberg:2012}, and \cite{sal:whi:gol:mur12}, respectively, published comprehensive survey articles discussing statistical approaches, challenges and developments in network data analysis. We also refer to the monograph of \cite{Kolaczyk:2009} for a comprehensive introduction to the field. \\
In this paper we consider networks represented as a $n \times n$ dimensional adjacency matrix $\bY$, where the element $Y_{ij}=1$, if an edge exists between vertex $i$ and vertex $j$ and $Y_{ij} = 0$ otherwise, with $i, j \in \{1, \ldots, n \}$ and $i \neq j$, that is there is no connection from a vertex to itself. With $n$ we denote the number of vertices in the network and for simplicity we assume undirected edges, that is $Y_{ij} = Y_{ji}$. Therefore, the matrix $\bY$ is symmetric and for simplicity it is sufficient to consider the upper triangle of $\bY$ only, that is $Y_{ij}, j > i$. Our approach equally applies to non-symmetric adjacency matrices corresponding to directed graphs. A concrete realisation of $\bY$ is denoted with $\by$. \vspace{6pt}\\
With respect to the available statistical models for modelling cross-sectional network data one may roughly distinguish between two strands, (a) models which explain the existence of an edge purely with external nodal covariates or random effects and (b) models where the existence of an edge also depends on the local network structure. The first strand of models is phrased as $p_1$ and $p_2$ models tracing back to \cite{HollandLeinhardt:1981}. Specifically, in the $p_1$ model we set 
\begin{align}\label{eq:p1}
\text{logit}\left[ \Prob(Y_{ij} = 1) \right] &= \log \left\lbrace \frac{ \Prob(Y_{ij} = 1) }{ 1 - \Prob(Y_{ij} = 1) } \right\rbrace = \alpha_i + \alpha_j + \boldsymbol{z}_{ij}^t \boldsymbol{\beta}
\end{align}  
where $\boldsymbol{z}_{ij}$ denotes a set of covariates relating to the vertices $i$ and $j$ and $\alpha_i$ and $\alpha_j$ are nodal effects, here assuming undirected edges. Since the number of parameters increases with increasing network size $n$, \cite{Duijn-etal:2004} proposed to replace the $\alpha$ parameters in (\ref{eq:p1}) by random effects, see also \cite{Zijlstra-etal:2006}. This yields the $p_2$ model 
\begin{align}\label{eq:p2}
\text{logit}\left[ \Prob(Y_{ij} = 1|\vecphi) \right] &= \phi_i + \phi_j + \boldsymbol{z}_{ij}^t \beta, \\
\vecphi = (\phi_1,\ldots, \phi_n)^t &\sim N(0, \sigma_\phi^2 I_n) \notag
\end{align}
with $I_n$ as $n$ dimensional unit matrix. A general principle with this approach is that vertices (or actors in the network, respectively) are not considered as homogeneous but heterogeneous, though their heterogeneity is not observable but latent and expressed in the node specific random effects $\phi_i,\ i = 1, \ldots, n$.\vspace{6pt}\\
Both, the $p_1$ and the $p_2$ model lie within the classical generalized linear (mixed) model framework which allows estimation using standard statistical software. The $p_2$ models also allow for Bayesian estimation approaches, see for example \cite{GillSwartz:2004}.\vspace{6pt}\\
The second strand in statistical network modelling is based on the so called Exponential Random Graph Model (ERGM) proposed by \cite{FrankStrauss:1986}. Here we model directly the network using the likelihood function 
\begin{align}\label{eq:ergm}
\Prob ( \bY = \by | \vectheta ) &= f( \by | \vectheta ) = \frac{ q_{\vectheta} ( \by ) }{ \kappa( \vectheta ) } = 
\frac{ \exp \left\{ \vectheta^t s( \by ) \right\} }{ \kappa( \vectheta ) }
\end{align}
where $\vectheta = (\theta_1, \ldots, \theta_p)^t$ is the vector of model parameters and $s(\by)$ is a vector of sufficient network statistics like the number of edges or two-stars in a network, see for example \cite{Snijders-etal:2006}. In equation (\ref{eq:ergm}) the term $\kappa(\vectheta)$ denotes the normalizing constant, that is
\[
\kappa(\vectheta) = \sum\limits_{\by \in \mathcal{Y}} \exp\left\{\vectheta^t s(\by) \right\}
\]
and is accordingly the sum over $2^{\binom{n}{2}}$ potential undirected graphs and therefore numerically intractable, except for very small graphs. Early fitting approaches are based on the pseudolikelihood idea proposed by \cite{StraussIkeda:1990}. More advanced are MCMC based routines proposed by \cite{HunterHandcock:2006} based on the work of \cite{GeyerThompson:1992}. A fully Bayesian approach to estimate ERGMs has been developed by \cite{CaimoFriel:2011}.\vspace{6pt}\\
Model (\ref{eq:ergm}) allows for a conditional interpretation by focusing on the occurrence of a single edge between two nodes. To be specific we obtain 
\begin{align}\label{eq:ergm-cond}
\text{logit}\left[ \Prob\bigl(Y_{ij}=1 | Y_{kl}, (k,l) \neq (i,j); \vectheta\bigr) \right]
&= \vectheta^t\ s_{ij}(\by),
\end{align} 
where $s_{ij}(\by)$ denotes the vector of so called change statistics
\[
s_{ij}(\by) = s \bigl(y_{ij} = 1,y_{kl}, (k,l) \neq (i,j)\bigr) -
s\bigl(y_{ij}=0,y_{kl},(k,l) \neq (i,j)\bigr).
\]
We refer to \cite{RobinsPattison-etal:2007} and \cite{RobinsSnijders-etal:2007} for a deeper discussion of Exponential Random Graph Models.\vspace{6pt}\\
Contrasting equation (\ref{eq:ergm-cond}) with the $p_1$ and $p_2$ model given in equations (\ref{eq:p1}) and (\ref{eq:p2}) it becomes obvious that the ERGM in contrast to the $p_1$ and $p_2$ models take the network structure into account while considering the nodes to be homogeneous. When modelling network data this means that all possible heterogeneity in the network nodes (that is the actors in the network) is included as covariates in the model and influence the (global) structure of the network. Since homogeneity of the nodes have led from $p_1$ to $p_2$ models, we want to pursue the same modelling exercise by allowing for latent node specific heterogeneity in Exponential Random Graph Models. To do so, we combine the $p_2$ model (\ref{eq:p2}) with the ERGM (\ref{eq:ergm-cond}) towards 
\begin{align}\label{eq:ergm-p2}
\text{logit}\left[ \Prob\bigl(Y_{ij}=1 | Y_{kl}, (k,l) \neq (i,j); \vectheta, \phi_i, \phi_j \bigr) \right]
&= \vectheta^t\ s_{ij}(\by) + \phi_i + \phi_j
\end{align} 
with $\vecphi = (\phi_1, \ldots, \phi_n)^t$ and $\phi_i \stackrel{\text{i.i.d.}}{\sim} N(\mu_\phi,\sigma_\phi^2),\ i = 1, \ldots, n$. The parameter $\mu_\phi$ captures the average propensity in the network for forming a tie. In terms of the likelihood function for the whole network we obtain from \eqref{eq:ergm-p2}
\begin{align}\label{eq:ergm-p2-lik}
\Prob ( \bY = \by | \vectheta, \vecphi ) &= f(\by|\vectheta, \vecphi) = \frac{ q_{\vectheta, \vecphi} ( \by ) }{ \kappa( \vectheta, \vecphi ) } = 
\frac{ \exp \left\{ \vectheta^t s( \by )+ \vecphi^t t(\by) \right\} }{ \kappa( \vectheta, \vecphi ) },
\end{align}
where $t(\by)$ contains the degree statistics of the $n$ vertices, i.e. $t_i(\by) = \sum\limits_{j = 1}^{n} y_{ij},\ \text{for } i = 1, \ldots, n$.
That is we fit an Exponential Random Graph Model with random, node specific effects accounting for heterogeneity. The model in equations (\ref{eq:ergm-p2}) and (\ref{eq:ergm-p2-lik}) falls in the general class of Exponential-family Random Network Models proposed by \cite{FellowsHandcock:2012} but unlike their model we treat the node specific effect as latent and we pursue a fully Bayesian estimation. We also refer to \cite{KrivitskyHandcock:2009} who propose a model with actor specific random effects based on a latent cluster model. The authors also propose node specific random effects. We follow this line and give further interpretability of the effects. A central issue in model extensions is the question of model selection.  
We emphasize this point in the paper by comparing models with and without nodal effects using the Bayes factor as model selection criterion. However, calculation of the Bayes factor suffers from the above mentioned problem in Exponential Random Graph Models in that the normalization constant $\kappa(\cdot)$ is numerically infeasible. We therefore propose an approximate calculation of the Bayes factor and show in a simulation study its usability for model selection. 
\vspace{6pt}\\
For estimation and model selection of model \eqref{eq:ergm-p2-lik} we extend the fully Bayesian approach from \cite{CaimoFriel:2011}. The developed estimation routine is based on the numerical work of \cite{CaimoFriel:2014} with their \textsf{R} \citep{RCore:2014} package {\tt Bergm} (see http://cran.r-project.org/web/packages/Bergm). Our algorithms for model fitting and selection will be included in the {\tt Bergm} package.\vspace{6pt}\\
The paper is organized as follows. In Section \ref{sec:Bayesian} we derive a fully Bayesian formulation of the model. This is followed by a detailed description of the MCMC based estimation routine. Section \ref{sec:Selection} deals with the issue of model selection using Bayes factors. Two data examples and some simulation results are presented Section \ref{sec:Examples}. Finally Section \ref{sec:Discussion} concludes with a discussion. 

\section{Bayesian model formulation and Estimation}
\label{sec:Bayesian}

\begin{figure}[h!]\centering
\includegraphics[scale=1.2]{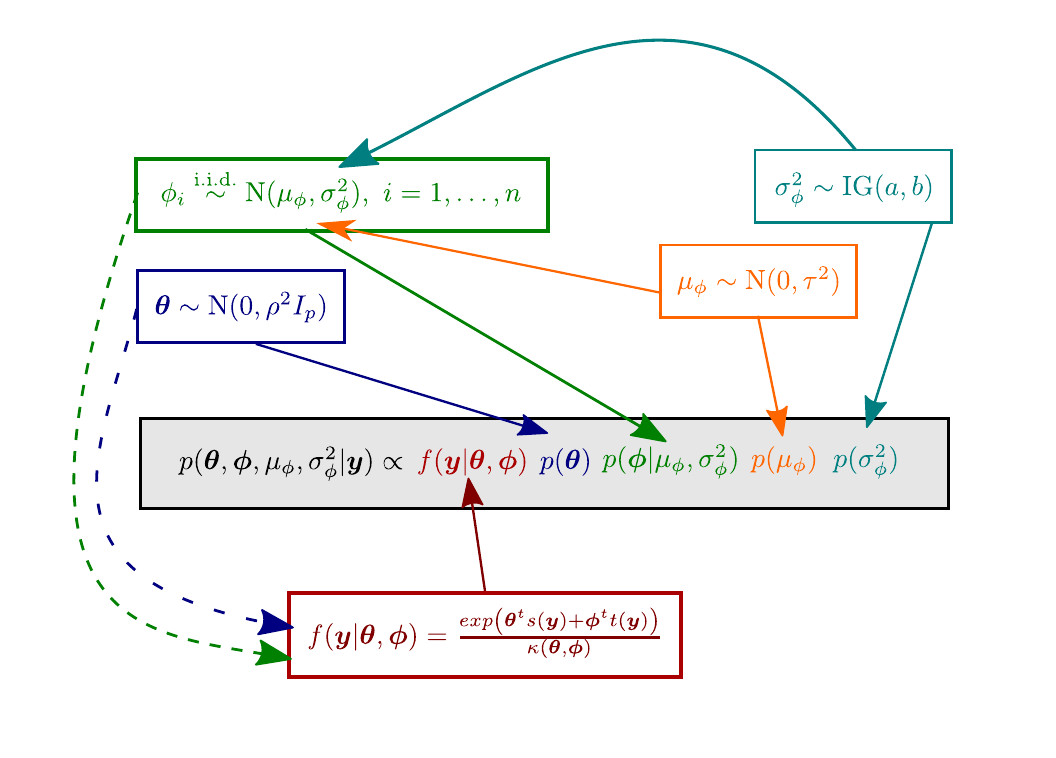}
\caption{\label{pic:overview_dist} Overview of the Bayesian model formulation for the Exponential Random Graph Model with nodal random effects.}
\end{figure}

Before proposing a fully Bayesian formulation for model \eqref{eq:ergm-p2-lik} bear in mind that the normalizing constant $\kappa(\vectheta,\vecphi)$ is numerically infeasible to calculate except for small networks so that numerically demanding simulation based fitting routines need to be employed. We follow a fully Bayesian approach by imposing a prior distribution on $\vectheta$. The posterior of interest for the Bayesian Exponential Random Graph Model with nodal random effects in (\ref{eq:ergm-p2-lik}) then becomes 
\begin{align}\label{eq:bergm_post_full}
p(\vectheta, \vecphi, \mu_\phi, \sigma^2_\phi|\by) &= 
\frac{f(\by|\vectheta,\vecphi) p(\vectheta) p(\vecphi|\mu_\phi, \sigma^2_\phi) p(\mu_\phi) p(\sigma^2_\phi)}{p(\by)} ,
\end{align}
where $p(\vectheta)$ is the prior distribution of $\vectheta$ and $p(\vecphi|\mu_\phi, \sigma^2_\phi)$ the prior for the random nodal effects $\vecphi$. We assume the nodal effects to be independent and identically normally distributed, that is
\[
\phi_i \sim \pnorm (\mu_\phi, \sigma^2_\phi), \quad \text{for } i = 1, \dots, n
\]
and accordingly we use $\vectheta \sim \pnorm (0, \rho^2 I_p)$, with $I_p$ denoting the $p$-dimensional unity matrix and $\rho^2$ chosen such that the prior distribution is flat. For the hyper prior distribution $p(\mu_\phi)$ of the mean $\mu_\phi$ we assume a normal distribution centred at 0, that is
\[
\mu_\phi \sim \pnorm (0, \tau^2).
\]
The hyper prior $p(\sigma^2_\phi)$ of the variance $\sigma^2_\phi$ is assumed to be an inverse gamma distribution, that is
\[
\sigma^2_\phi \sim \pinvgamma(a,b).
\]
Finally, the parameters $\tau^2$, $a$ and $b$ are all constants and chosen in a way that results in flat hyper prior distributions.  Figure \ref{pic:overview_dist} illustrates this Bayesian model formulation. 

It is important to note, that the posterior distribution in (\ref{eq:bergm_post_full}) is so-called doubly-intractable. This is because, firstly, it is not possible to evaluate the posterior density (\ref{eq:bergm_post_full}) due to $p(\by)$, the marginal likelihood or evidence, being intractable. Secondly, it is also numerically infeasible to calculate the normalizing constant $ \kappa( \vectheta, \vecphi )$ in the likelihood $f(\by|\vectheta,\vecphi)$ except for very small network graphs. Similar to the algorithm proposed by \cite{CaimoFriel:2011} we use the so-called exchange algorithm from \cite{Murray-etal:2006} to draw samples from the posterior distribution of interest. Let therefore $\vecgamma = (\vectheta,\vecphi)$ denote the entire parameter vector of the ERGM. 
Instead of drawing directly from (\ref{eq:bergm_post_full}), we sample from the augmented distribution
\begin{align}\label{bergm_post_augmented}
& p(\vecgamma', \by', \vecgamma, \mu_\phi, \sigma^2_\phi|\by) \propto \notag\\
& \quad 
f( \by | \vecgamma ) p(\vecgamma|\mu_\phi, \sigma^2_\phi) p(\mu_\phi) p(\sigma^2_\phi) h(\vecgamma' | \vecgamma) f( \by' | \vecgamma' ),
\end{align}
where $h( \cdot | \cdot )$ is a proposal function, to be specified later. This proposal provides $\vecgamma' = (\vectheta',\vecphi')$
as new candidate values for $\vectheta$ and $\vecphi$, respectively, and based on $\vecgamma'$ we can simulate $\by'$ as an auxiliary network. The proposal is accepted with probability 
\begin{align}
\label{eq:alpha}
\alpha &= \min \left( 1, \frac{ q_{\vecgamma} ( \by' ) p( \vecgamma' ) h( \vecgamma | \vecgamma' ) q_{ \vecgamma' } ( \by ) }{ q_{ \vecgamma } ( \by ) p( \vecgamma ) h( \vecgamma' | \vecgamma ) q_{ \vecgamma' } ( \by' ) } \times \frac{ \kappa(\vecgamma) \kappa (\vecgamma') }{\kappa(\vecgamma) \kappa (\vecgamma')} \right)
\end{align}
where $p(\vecgamma) = p(\vectheta) \cdot p( \vecphi | \mu_\phi, \sigma^2_\phi ).$ Note that in \eqref{eq:alpha} the normalizing constants cancel out so that \eqref{eq:alpha} is in principle easy to calculate. Though the algorithm is in this form a direct extension of the BERGM algorithm in \cite{CaimoFriel:2011} it is advisable to separate the proposals of $\vectheta$ and $\vecphi$ to achieve higher acceptance rates. This is described in the following algorithmic steps.
In detail, our algorithm works as follows:
\begin{framed}
\textbf{Algorithm 1:} Fit BERGM with nodal random effects
\hrule
\begin{enumerate}
\renewcommand{\labelenumi}{\textit{Step \arabic{enumi}:}}
\item Gibbs update of $(\vectheta', \by')$: 
	\begin{enumerate}[(i)]
		\item Draw $\vectheta' \sim h(\cdot | \vectheta)$.
		\item Draw $\by' \sim p(\cdot | \vectheta', \vecphi)$.
		\item Propose to move from $\vectheta$ to $\vectheta'$ with probability
\begin{align*}
\alpha &= \min \left( 1, \frac{ q_{ \vectheta, \vecphi } ( \by' ) p( \vectheta' ) h( \vectheta | \vectheta' ) q_{ \vectheta', \vecphi } ( \by ) }{ q_{ \vectheta, \vecphi } ( \by ) p( \vectheta ) h( \vectheta' | \vectheta ) q_{ \vectheta', \vecphi } ( \by' ) } \times \frac{ \kappa(\vectheta, \vecphi) \kappa (\vectheta', \vecphi) }{\kappa(\vectheta, \vecphi) \kappa (\vectheta', \vecphi)} \right).
\end{align*}
	\end{enumerate}
\item Gibbs update of $(\vecphi', \by')$: 
	\begin{enumerate}[(i)]
		\item Draw $\vecphi' \sim g(\cdot | \vecphi)$.
		\item Draw $\by' \sim p(\cdot | \vectheta, \vecphi')$.
		\item Propose to move from $\vecphi$ to $\vecphi'$ with probability
\begin{align*}
\alpha &= \min \left( 1, \frac{ q_{ \vectheta, \vecphi } ( \by' ) p( \vecphi' | \mu_\phi, \sigma^2_\phi ) g( \vecphi | \vecphi' ) q_{ \vectheta, \vecphi' } ( \by ) }{ q_{ \vectheta, \vecphi } ( \by ) p( \vecphi | \mu_\phi, \sigma^2_\phi ) g( \vecphi' | \vecphi ) q_{ \vectheta, \vecphi' } ( \by' ) } \times \frac{ \kappa(\vectheta, \vecphi) \kappa (\vectheta, \vecphi') }{\kappa(\vectheta, \vecphi) \kappa (\vectheta, \vecphi')} \right).
\end{align*}
	\end{enumerate}
\item Metropolis-Hastings update of $\mu_\phi$: 
\newline Draw proposal $\mu_\phi'$ from $k(\cdot | \mu_\phi)$ and accept the proposed value with probability $\alpha = \min \left( 1, \frac{ p(\vecphi | \mu_\phi, \sigma^2_\phi) p(\mu_\phi) }{ p(\vecphi | \mu_\phi', \sigma^2_\phi) p(\mu_\phi') } \right)$.
\item Metropolis-Hastings update of $\sigma^2_\phi$:
\newline Draw proposal ${\sigma_\phi^2}'$ from $l(\cdot | \sigma^2_\phi)$ and accept the proposed value with probability $\alpha = \min \left( 1, \frac{ p(\vecphi | \mu_\phi, \sigma^2_\phi) p(\sigma^2_\phi) }{ p(\vecphi | \mu_\phi, {\sigma_\phi^2}') p({\sigma_\phi^2}') } \right)$.
\end{enumerate}
Start again with \textit{Step 1} until the maximum number of iterations is reached.
\end{framed}

It is easy to see that there is no necessity to compute the normalizing constants $\kappa(\cdot)$, because they cancel out when calculating the acceptance probabilities in the first two steps of the algorithm. The current implementation of the algorithm uses single-site updates for the update of $\vecphi$, that is each $\phi_i,\ i = 1, \ldots, n$ is updated in turn while all other values are kept constant. This leads to reasonable acceptance probabilities for the Markov chain. \\
The default choices for the proposal functions $h(\cdot | \cdot)$, $g(\cdot | \cdot)$ and $k(\cdot | \cdot)$ are normal distributions centred at the current parameter value, for $l(\cdot | \cdot)$ we use a uniform distribution, which is symmetric around the current value of $\sigma^2_\phi$ and truncated at zero to avoid negative proposals for the variance parameter. \\
The draws of the auxiliary network $\by'$ in the component of steps 1 and 3 are realised using the ``tie no tie'' sampler from the \texttt{ergm} package \citep{HunterHandcock:2008}, which is a simple Gibbs sampler. Although this auxiliary Gibbs sampler does not yield an exact draw $\by'$, \cite{Everitt:2012} has shown, under some assumptions, that the resulting approximate exchange algorithm converges to the target distribution as the number of auxiliary draws tends to infinity. As a practical result he points out that for the number of auxiliary iterations it is often sufficient to use roughly the number of possible ties in the network. 

\section{Model Selection}
\label{sec:Selection}

Model Selection is an important, often neglected issue in network data analysis. We put special emphasis on this task here and propose the Bayes factor suitable for model selection. One of the interesting questions in our model is, if we are able to distinguish the three following model generating processes:
\begin{enumerate}
\item Nodal random effects only, i.e. the $p_2$ model,
\item Structural effects only, i.e. the standard ERGM, and
\item ERGM in combination with nodal random effects.
\end{enumerate}
This question results in the problem of model selection. The data examples in Section \ref{sec:Data} illustrate this issue. \\
Classical Bayesian tools for model comparison such as the deviance information criterion (DIC) as suggested by \cite{Spiegelhalter-etal:2002} are not directly available, again due to the intractability of the normalizing constant of the likelihood in model equation (\ref{eq:ergm-p2-lik}). \\
Computing Bayes factors for model choice using reversible jump Markov Chain Monte Carlo for Bayesian Exponential Random Graph Models as done by \cite{CaimoFriel:2013sel} is not an option for our model. This approach would be possible in general, but very time consuming from a computational point of view. \vspace{6pt}\\
We suggest the following strategy for deciding whether to include nodal random effects into the model or not. The goal is to calculate a Bayes factor for two competing models \citep{KassRaftery:1995}. First we fit the two Exponential Random Graph Models, one with edges and non-random effects only, notated as model $m_1$ with coefficients $\vectheta'$, and the second one with nodal random effects instead of the edges term, labelled as model $m_2$ with coefficients $\vectheta$ and $\vecphi$. Note that these two models are nested.\footnote{For the approach presented here, especially for the path sampling, we need the models to be nested. In general it would be possible to extend the approach to non-nested models as well.} \\
Following Bayes theorem the so-called evidence for each model can be calculated using 
\begin{align}\label{eq:evm1}
p( \by | m_1 ) &= \frac{f ( \by | \vectheta' ) p ( \vectheta' ) }{ p ( \vectheta' | \by ) }, \quad \forall\ \vectheta',
\end{align} 
for model $m_1$, and 
\begin{align}\label{eq:evm2}
p( \by | m_2 ) &= \frac{f ( \by | \vectheta, \vecphi, \mu_\phi, \sigma^2_\phi ) p ( \vectheta ) p ( \vecphi | \mu_\phi, \sigma^2_\phi ) p (\mu_\phi) p ( \sigma^2_\phi ) }{ p ( \vectheta, \vecphi, \mu_\phi, \sigma^2_\phi | \by ) }, \quad \forall\ \vectheta, \vecphi, \mu_\phi, \sigma^2_\phi, \notag\\
&= \frac{f ( \by | \vectheta, \mu_\phi, \sigma^2_\phi ) p ( \vectheta ) p (\mu_\phi) p ( \sigma^2_\phi ) }{ p ( \vectheta, \mu_\phi, \sigma^2_\phi | \by ) }, \quad \forall\ \vectheta, \mu_\phi, \sigma^2_\phi
\end{align} 
for model $m_2$. \\
The term $f( \by | \vectheta, \mu_\phi, \sigma^2_\phi )$ denotes the marginal likelihood from model $m_2$, where the random effects $\vecphi$ have been marginalized, i.e.
\begin{align}\label{eq:marlik}
f( \by | \vectheta, \mu_\phi, \sigma^2_\phi )
&= \displaystyle\int \frac{ \exp\left\{ \vectheta^t s( \by ) + \vecphi^t t( \by ) \right\} }{ \kappa( \vectheta, \vecphi ) } \cdot p( \vecphi | \mu_\phi, \sigma^2_\phi ) \ \mbox{d} \vecphi \notag\\
&\approx \frac{ \exp\left\{ \vectheta^t s( \by ) \right\} }{ \kappa( \vectheta, \widehat{\vecphi} ) } \widehat{f}_\text{Laplace}( \by | \widehat{\vecphi}, \mu_\phi, \sigma^2_\phi ).
\end{align}
The approximation in equation (\ref{eq:marlik}) is achieved using a Laplace approximation around the point $\widehat{\vecphi}$. Details of this approximation are given in Section \ref{app:sec:laplace} of the appendix. \\
The Bayes factor of model $m_2$ against model $m_1$ is then defined as the ratio of (\ref{eq:evm2}) and  (\ref{eq:evm1}), i.e.
\begin{align}\label{eq:bf}
\text{BF}_{21} = 
\frac{ p( \by | m_2 ) }{ p( \by | m_1 ) } 
&= \frac{ f( \by | \vectheta, \mu_\phi, \sigma^2_\phi ) }{ f( \by | \vectheta' ) } \cdot \frac{ p( \vectheta ) p( \mu_\phi ) p( \sigma^2_\phi ) }{ p( \vectheta' ) } \cdot \frac{ p( \vectheta' | \by ) }{ p( \vectheta, \mu_\phi, \sigma^2_\phi | \by ) }.
\end{align}
Applying the approximation from equation (\ref{eq:marlik}) to (\ref{eq:bf}), and plugging in estimates for the posterior densities
\begin{align}\label{eq:postest}
p( \vectheta' | \by ) \approx \widehat{p}( \vectheta' | \by ) 
\quad &\text{and} \quad 
p( \vectheta, \mu_\phi, \sigma^2_\phi | \by ) \approx \widehat{p}( \vectheta, \mu_\phi, \sigma^2_\phi | \by )
\end{align}
leads to 
\begin{align}\label{eq:bf_ext}
\text{BF}_{21}
&\approx \frac{ \exp\left\{ \vectheta^t s( \by ) \right\} \widehat{f}_\text{Laplace}( \by | \widehat{\vecphi}, \mu_\phi, \sigma^2_\phi ) }{ \exp\left\{ \vectheta'^t s'( \by ) \right\} } \cdot 
\frac{ \kappa( \vectheta' ) }{ \kappa( \vectheta, \widehat{\vecphi} ) } \cdot \frac{ p( \vectheta ) p( \mu_\phi ) p( \sigma^2_\phi ) }{ p( \vectheta' ) } \cdot \frac{ \widehat{p}( \vectheta' | \by ) }{ \widehat{p}( \vectheta, \mu_\phi, \sigma^2_\phi | \by ) }.
\end{align}
The ratio of the two normalizing constants $\kappa( \vectheta' )\ /\ \kappa( \vectheta, \widehat{\vecphi} )$ in (\ref{eq:bf_ext}) is estimated using a path sampling approach \citep{GelmanMeng:1998}, which is similarly used by \cite{CaimoFriel:2013sel}. Consider 
\[
\kappa( \vectheta(g), \vecphi(g) ),
\]
where 
\begin{align*}
\vectheta(g) 
&= (1 - g) \cdot \vectheta' + g \cdot \left[ \begin{array}{c}
0\\
\vectheta\\
\end{array} \right] \ \text{and} \\
\vecphi(g) 
&= g \cdot \vecphi
\end{align*}
for $g \in [0, 1]$. So by construction
\[
( \vectheta(0), \vecphi(0) ) = (\vectheta', \boldsymbol{0}) 
\quad \text{and} \quad
( \vectheta(1), \vecphi(1) ) = \left( \left[ \begin{array}{c}
0\\
\vectheta\\
\end{array} \right], \vecphi \right). \footnote{Note that the additional $0$ entry is necessary because the mixed effects model contains no parameter for the edges statistic. The edge effect is captured in the mean value $\mu_\phi$ of the nodal random effects. The missing edges statistics is also the difference between $s'(\by)$ and $s(\by)$.}
\]
Then thermodynamic integration (or so-called path sampling) can be used to estimate 
\begin{align*}
\log \left\{ \frac{\kappa( \vectheta' )}{ \kappa( \vectheta, \vecphi ) }\right\}
&= \displaystyle\int\limits_0^1 \E_{\bY | \vectheta(g), \vecphi(g)} \left[ \left( \vectheta' - \left[ \begin{array}{c}
0\\
\vectheta\\
\end{array} \right] \right)^t s'(\bY) + (- \vecphi)^t t(\bY) \right] \mbox{d} g.
\end{align*}
Consider discretising $g \in [0, 1]$ as $(g_0 = 0, \dots, g_i = \frac{i}{I}, \dots, g_I = 1)$. Then we approximate 
\begin{align*}
E_i 
&:= \E_{\bY | \vectheta(g_i), \vecphi(g_i)} \left[ \left( \vectheta' - \left[ \begin{array}{c}
0\\
\vectheta\\
\end{array} \right] \right)^t s'(\bY) + (- \vecphi)^t t(\bY) \right] \\
&\approx \frac{1}{N} \sum\limits_{j = 1}^N \left[ \left( \vectheta' - \left[ \begin{array}{c}
0\\
\vectheta\\
\end{array} \right] \right)^t s'(\by^{(j)}) + (- \vecphi)^t t(\by^{(j)}) \right],
\end{align*}
where the networks $\by^{(j)}$ are drawn from $f( \by | \vectheta(g_i), \vecphi(g_i) )$, for $j = 1, \dots, N$. Then we use a trapezoidal rule to numerically integrate
\begin{align*}
\log \left\{ \frac{\kappa( \vectheta' )}{ \kappa( \vectheta, \vecphi ) }\right\} 
&= \sum\limits_{i = 1}^{I - 1} (g_{i+1} - g_i) \cdot \left( \frac{E_{i+1} + E_i}{2}\right).
\end{align*}
The path sampling routine can easily be parallelised because the evaluations at the individual grid points of $g$ do not depend on each other.\\\vspace{3pt}
 
The Bayes factor in equation (\ref{eq:bf_ext}) is evaluated using the posterior mean values for the parameters $\vectheta$, $\vectheta'$, $\mu_\phi$ and also for $\widehat{\vecphi}$. For $\sigma^2_\phi$ we plug in the mean of the logarithmized values and transform it back onto the scale of $\sigma^2_\phi$, because the posterior density of $\sigma^2_\phi$ is not symmetric. \\
For reasons of simplicity the posterior density estimates (\ref{eq:postest}) are estimated assuming asymptotic normality, again using $\log(\sigma^2_\phi)$. For the data examples in the next section this assumption seems to be reasonable when looking at the plotted posterior density estimates. Furthermore, the individual contributions of the different components of the Bayes factor calculation suggest that at least in these cases the posterior density estimates play a minor part compared to the other components. If this assumption is violated this step in the algorithm can be changed.

\section{Examples}
\label{sec:Examples}

\subsection{Data Examples}
\label{sec:Data}

\subsubsection{Zachary's Karate Club Network}
\begin{figure}[htb!]\centering
\includegraphics[scale=0.8]{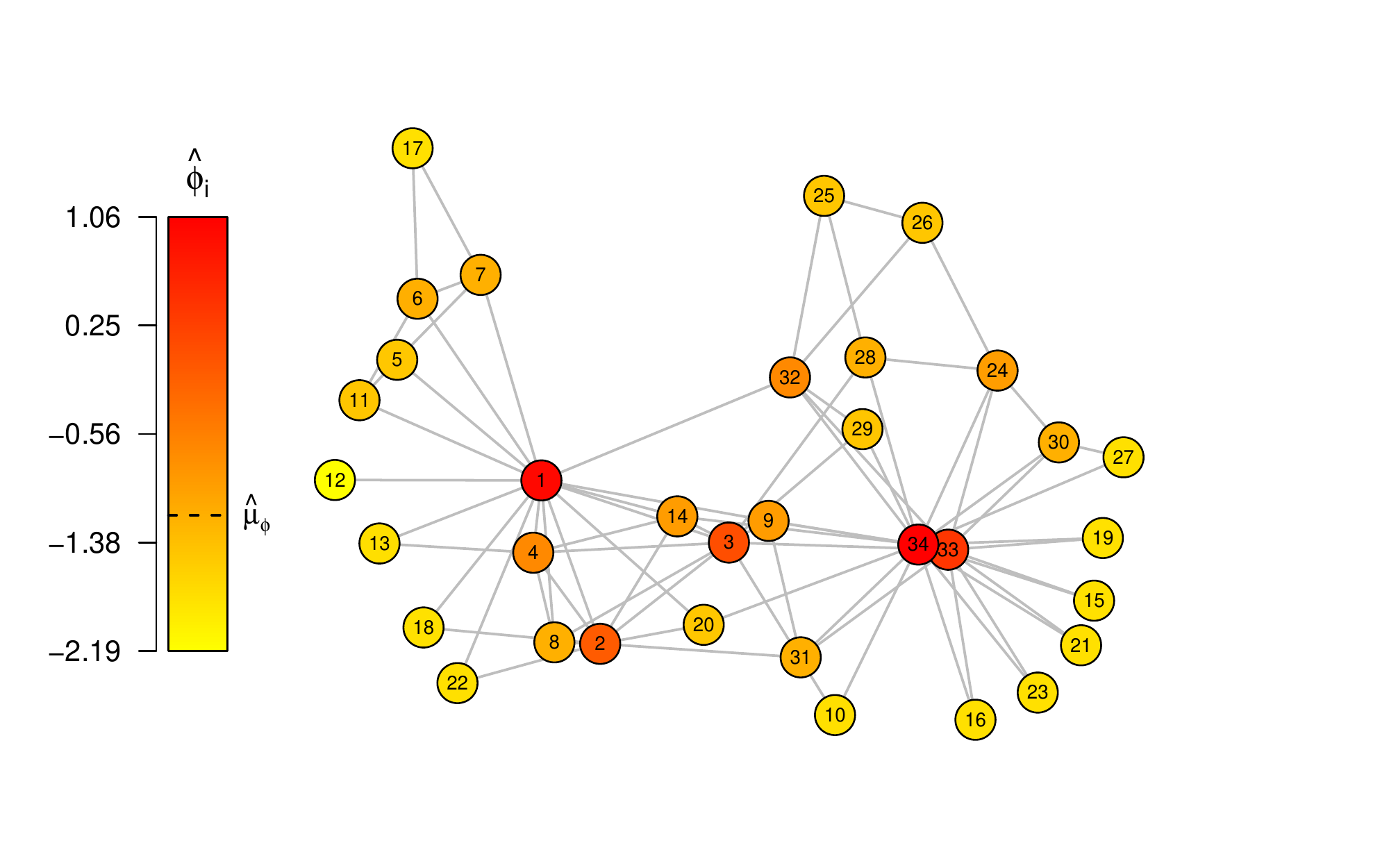}
\caption{Zachary's Karate Club Graph. Vertices are coloured by their estimated nodal effect $\widehat{\phi}_i$ (posterior mean), $i = 1, \ldots, n$. Vertices with a high nodal effect are darker in orange/red.}
\label{fig:zach_graph}
\end{figure}

As a first data example we employ Zachary's karate club network \citep{Zachary:1977} which is a very well known data set often used in network analysis. The undirected 34 node network represents the friendships among members of a university karate club. Figure \ref{fig:zach_graph} shows a plot of this network graph. It is evident that there are only some nodes with a very high degree (no. 1, 33, and 34) while the majority of the remaining vertices has only two to four links. If there are no additional nodal attributes available, that might explain some differences between the actors, like for example status in the club (trainer, student, etc.), the assumption of vertex homogeneity in a standard ERGM appears to be at least questionable.

We fitted two different models to the data: a standard ERGM with edges and triangles as sufficient statistics, and a model with nodal random effects and the triangle statistic. These two models are nested. \\
For the model fitting tasks we used the \texttt{Bergm} package \citep{CaimoFriel:2014} and our extension of the \texttt{Bergm} routines, respectively. With 1,000 burn-in iterations, 30,000 main iterations, and 3,000 auxiliary iterations for the network simulation in each MCMC step, the computation of the fixed model took about two minutes on a 2.1 Ghz processor, the mixed model needed about one hour and forty minutes. Using 3,000 auxiliary iterations should be large enough because we have 561 possible ties in the network. Again, we refer to the results of \cite{Everitt:2012}.\\
Table \ref{tab:zach_results} shows the resulting posterior estimates for both models. 

\begin{table}[htb!]\centering
\caption{\label{tab:zach_results} Model fitting results for the karate club data.}
\begin{tabular}{llcccc}
\hline\\[-1ex]
Model type & Parameter & Post. mean & Post. Sd. & Acceptance rate & Note \\[1ex]
\hline\hline\\[-1ex]
\multirow{2}{*}{fixed} & $\theta_\text{edges}$ & -2.32 & 0.16 & \multirow{2}{*}{0.43}&\\
& $\theta_\text{triangles}$ & 0.54 & 0.11 && \\[1ex]
\hline\\[-1ex]
\multirow{3}{*}{mixed} & $\mu_\phi$ & -1.17 & 0.22 & 0.26 &\\
& $\sigma^2_\phi$ & 1.05 & 0.58 & 0.54 & *\\
& $\theta_\text{triangles}$ & -0.04 & 0.21 & 0.09 & \\[1ex]
\hline\\[-1ex]
\multicolumn{6}{p{14cm}}{\footnotesize{* For $\sigma^2_\phi$ the posterior mean is calculated based on the logarithmized values and than transformed back to the scale of $\sigma^2_\phi$ (this leads to the geometric mean) due to the non-symmetric posterior density in this case.}}
\end{tabular}
\end{table}

\begin{figure}[htb!]\centering
  \includegraphics[scale=0.6]{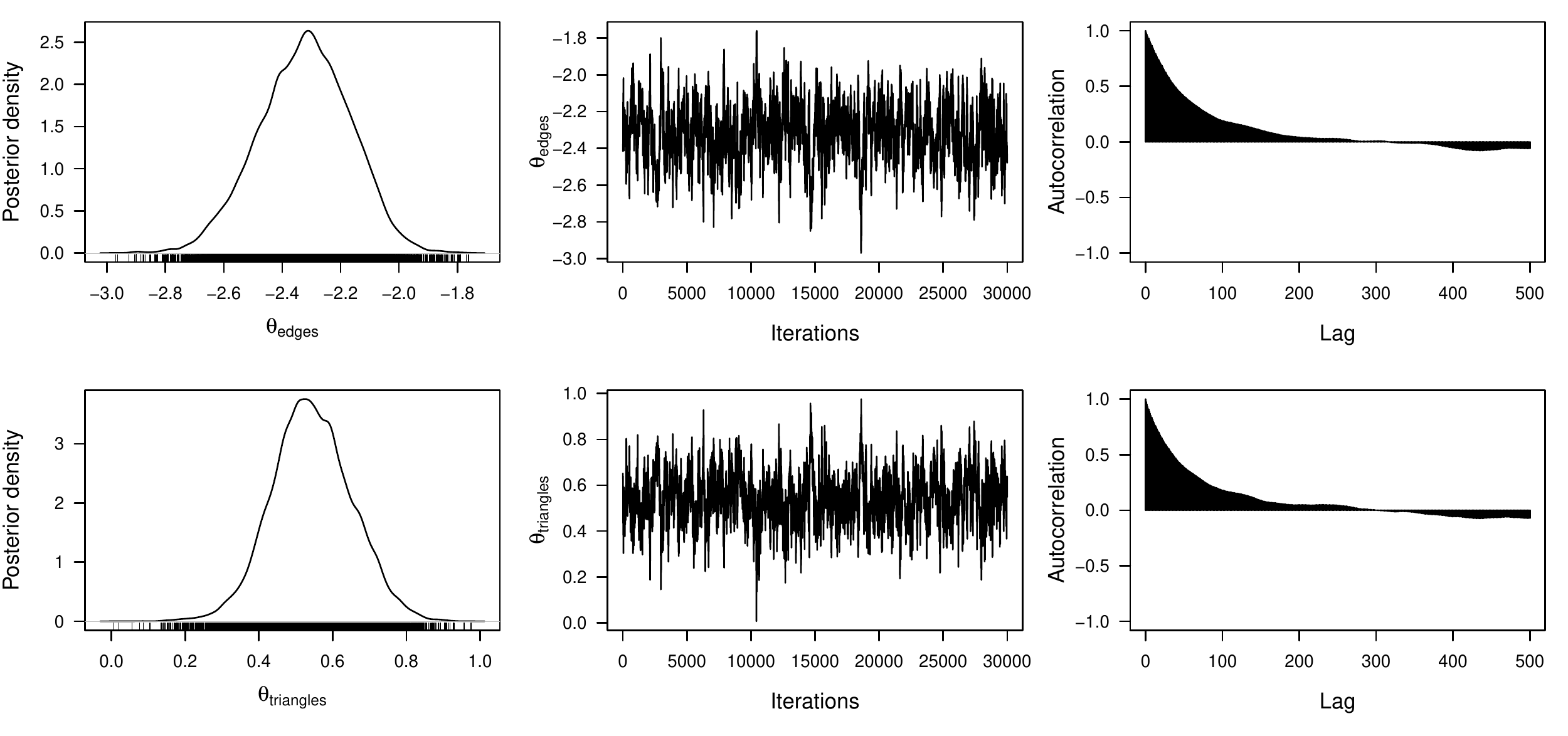} 
\caption{Posterior densities, trace plots, and autocorrelation for the fixed model with edges and triangular effect for the karate club data.}
\label{fig:zach_results_fixed}
\end{figure}

Figure \ref{fig:zach_results_fixed} shows the results for the fixed model with edges and triangular effect only. 

\begin{figure}[htb!]\centering
  \includegraphics[scale=0.6]{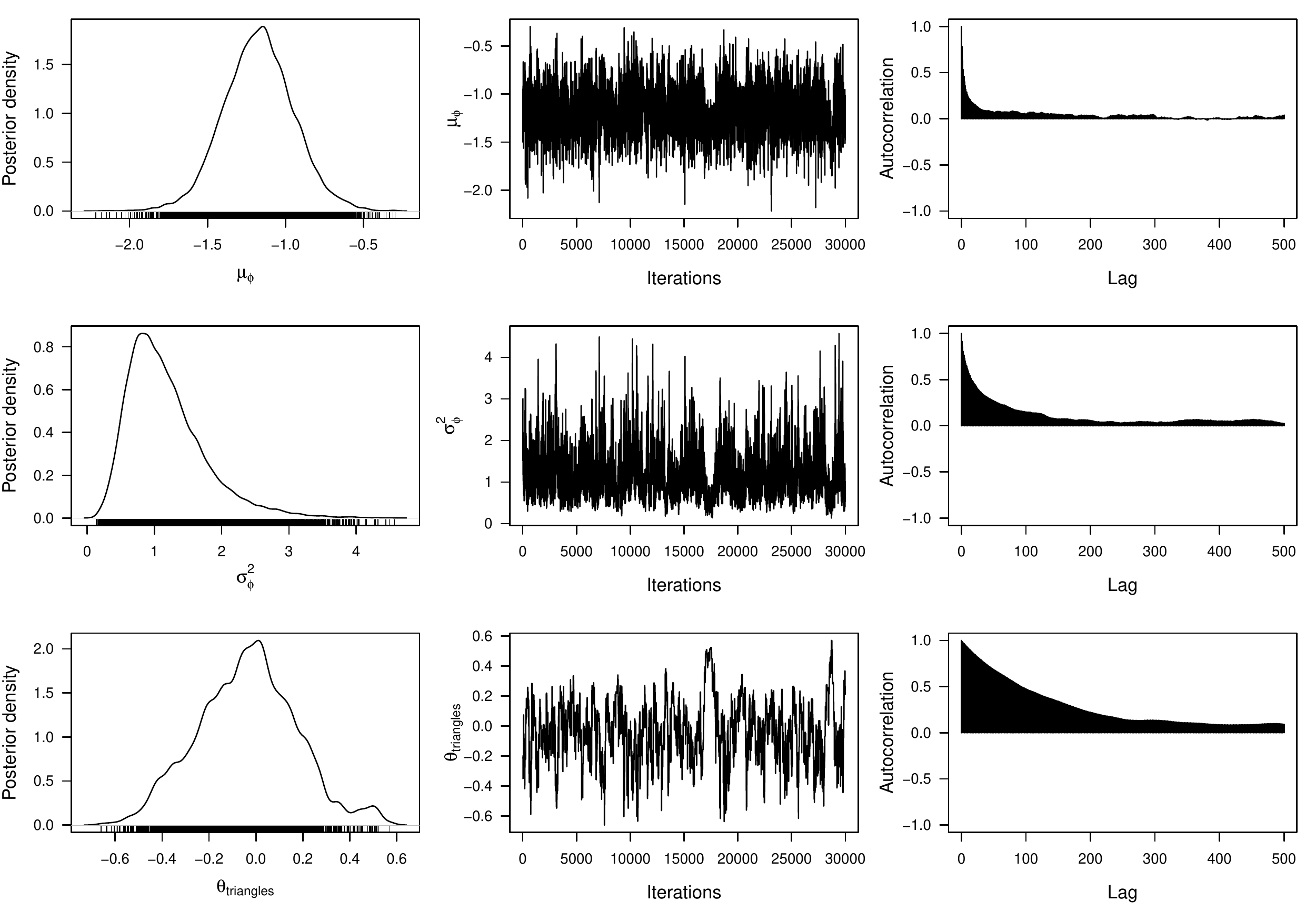} 
\caption{Posterior densities, trace plots, and autocorrelation for the mixed model with nodal random and triangular effects for the karate club data.}
\label{fig:zach_results_mixed}
\end{figure}

Figure \ref{fig:zach_results_mixed} shows the results for the mixed model with nodal random and triangular effects for the karate club data. 

\begin{figure}[htb!]\centering
\begin{tabular}{cc}
  \includegraphics[scale=0.5]{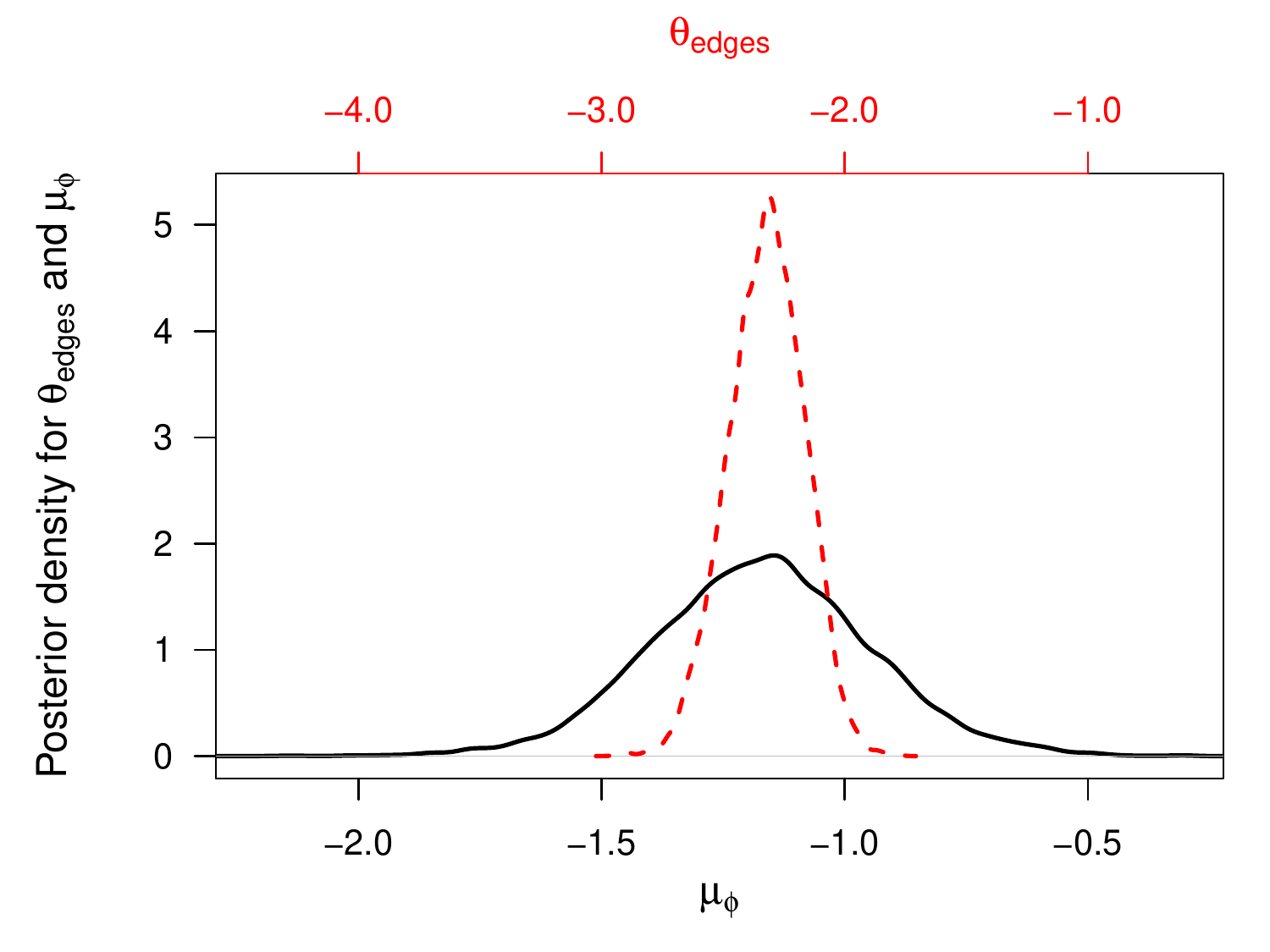} &   \includegraphics[scale=0.5]{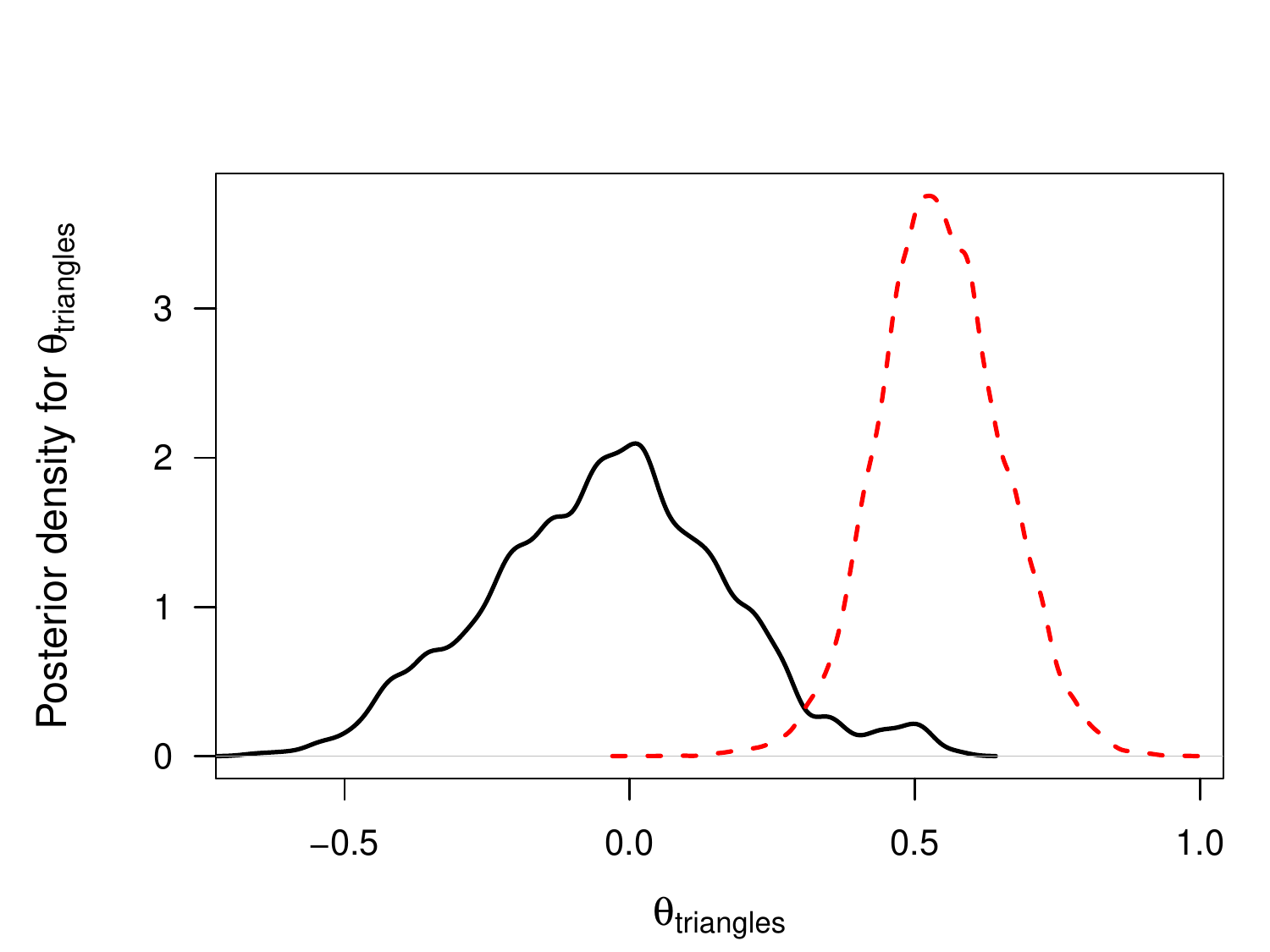} \\
\multicolumn{2}{c}{\includegraphics[scale=0.5]{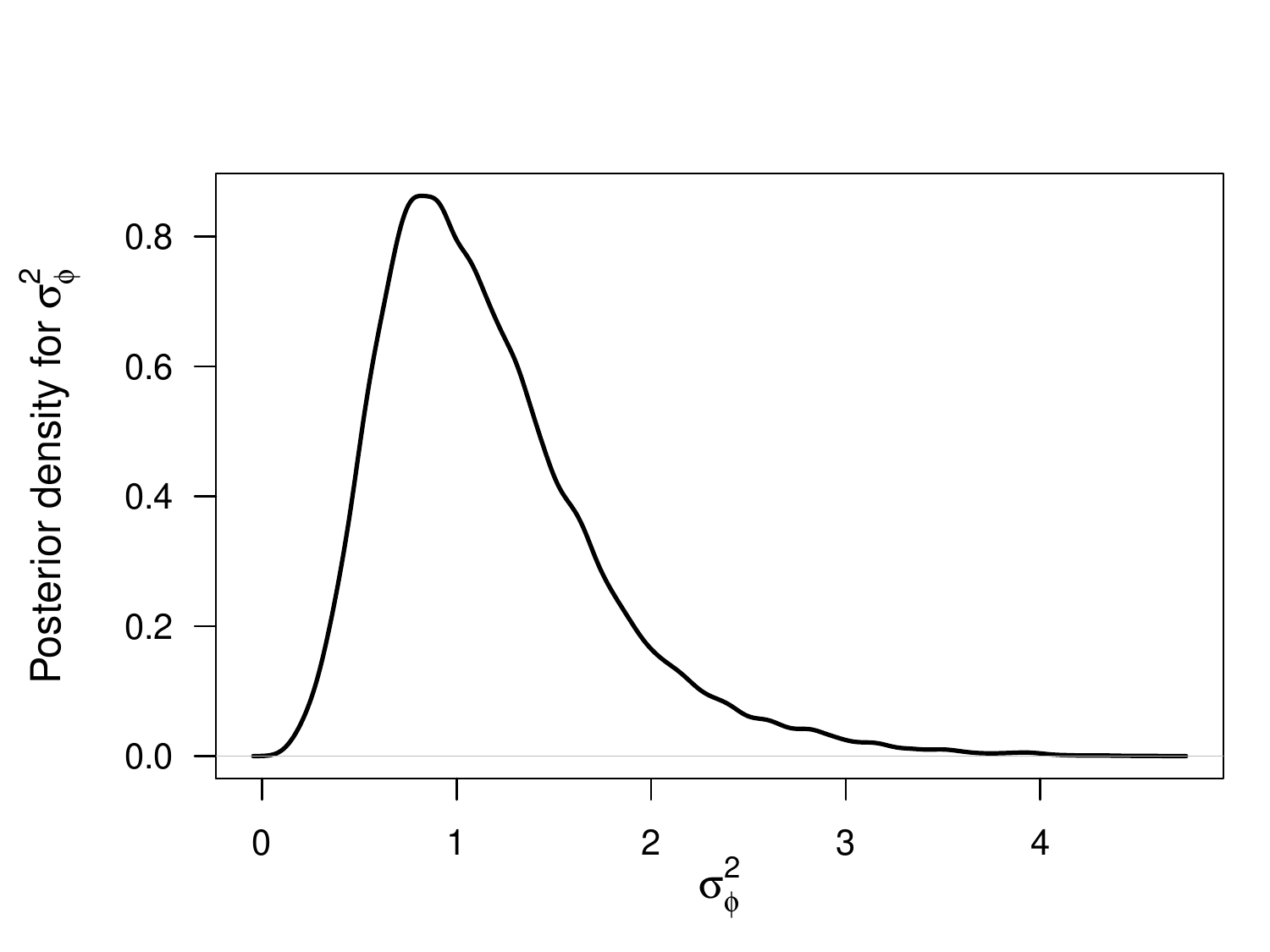} }\\
\end{tabular}
\caption{Posterior densities for the model with edges and triangular effect (red dashed lines) and nodal random and triangular effects (black solid lines) for the karate club data.}
\label{fig:zach_results}
\end{figure}

Figure \ref{fig:zach_results} shows estimates for the posterior densities for both models simultaneously. 

What is evident from the estimated posterior densities is the difference for the triangular effect in both models in the upper right plot of Figure \ref{fig:zach_results}. When not accounting for nodal heterogeneity this effect is clearly positive compared to the mixed model where the posterior support clearly comprises zero. For the parameter $\theta_\text{edges}$ associated with the edges statistic and $\mu_\phi$ in the mixed model there is no big difference between both models concerning the location of the posterior (when comparing $\theta_\text{edges}$ to $2 \cdot \mu_\phi$; note the different axis annotations). \\
The different effect of the triangular statistic in both models clearly illustrates the issue of model selection. After fitting the two competing models we computed a Bayes factor using the approach described in Section \ref{sec:Selection} to compare the model with nodal random effects to the one with structural effects only and tackle this issue. \\
The resulting estimated log Bayes factor is $453$, which is huge. As explained in the previous section, there is some randomness involved in the procedure.  Repeated calculation led to similarly huge values. This clearly indicates that the model with nodal random effects is preferable to the one without and is not surprising, because here a model with nodal heterogeneity appears much more realistic than one without. \\
Computing a single Bayes factor took about fourteen minutes using five 2.2 Ghz cores in parallel, with 10,000 iterations for the Laplace approximation, 1,000 grid points, 1,000 iterations at each point for the path sampling, and 3,000 iterations for each network simulation.\\

The vertices of the karate network in Figure \ref{fig:zach_graph} are coloured according to their estimated nodal effect $\widehat{\phi}_i$, $i = 1, \ldots, n$. As an estimate we use the corresponding posterior mean of each parameter $\phi_i$. Vertices with a high value are darker in orange/red. By using such a colouring scheme we are able to visualise the variation in the nodal effects. In addition, we can identify important nodes in the network based on the estimated nodal effects.

\subsubsection{European Parliament Members}

\begin{figure}[htb!]\centering
\includegraphics[scale=0.7]{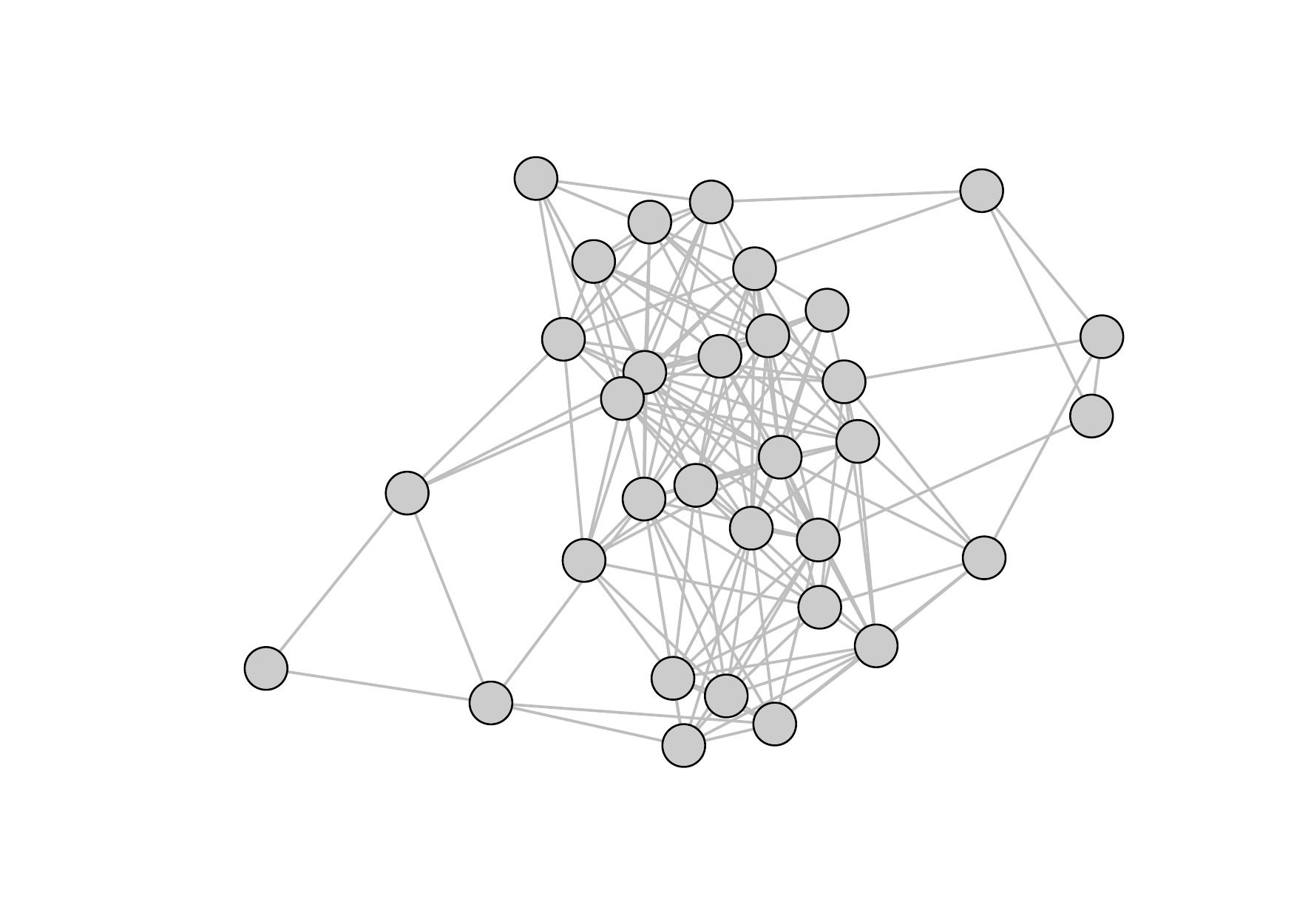}
\caption{Network of Dutch members of the European parliament during the 6th legislative period. Two members are linked if they have at least one committee membership in common.}
\label{fig:ep_nl_graph}
\end{figure}

The second data example consists of a network of members of the European parliament (MEP) in of the 6th legislative period. The complete network contains more than 900 vertices. We analyse a subset of the 32 members from the Netherlands. The induced subgraph is shown in Figure \ref{fig:ep_nl_graph}. A link between to MEPs exists if they have at least one committee membership in common. The data were provided by Paul W. Thurner \citep[see][]{Thurner-etal:2013}. This data example illustrates our model selection procedure. \\

\begin{figure}[htb!]\centering
  \includegraphics[scale=0.6]{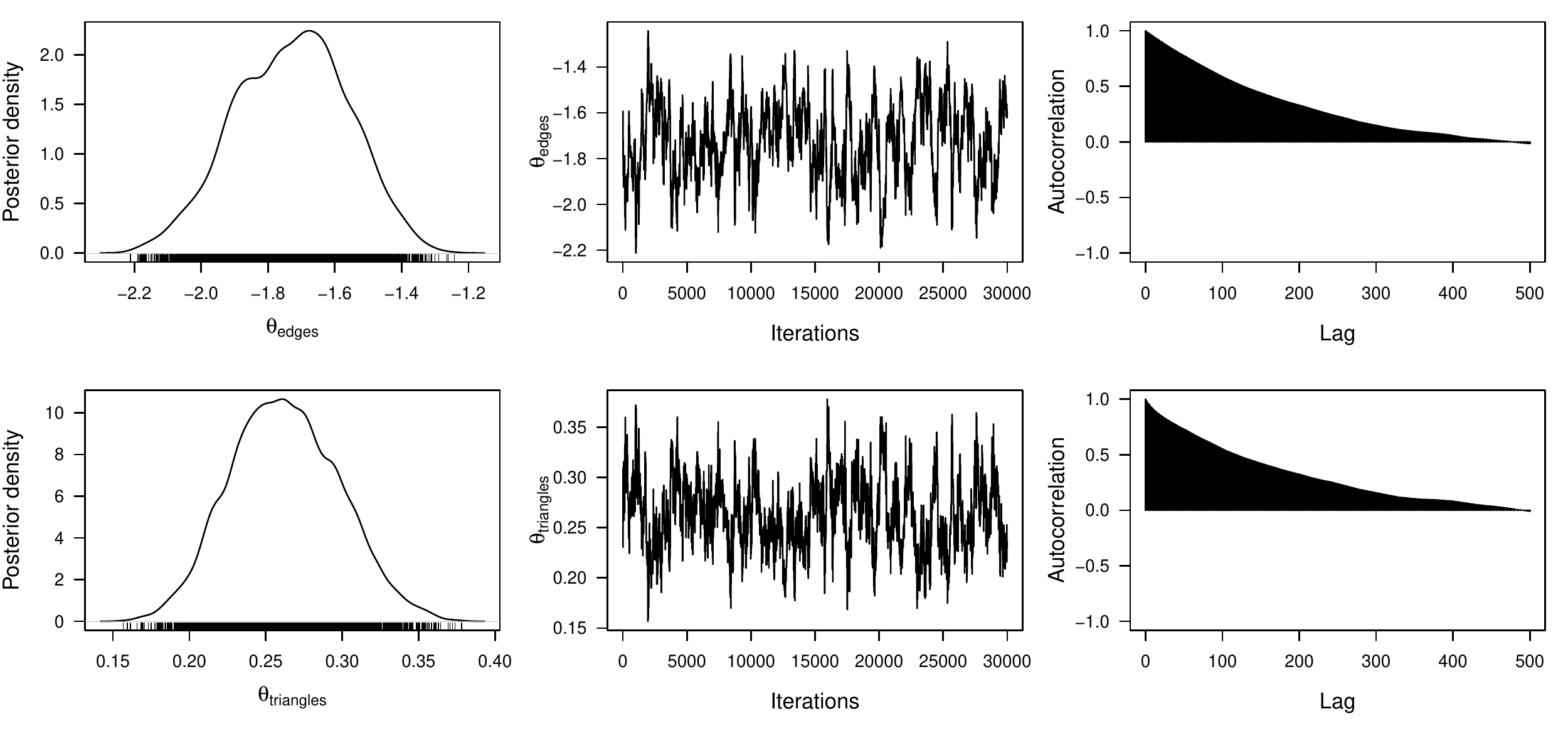} 
\caption{Posterior densities, trace plots, and autocorrelation for the fixed model with edges and triangular effect for the European parliament data.}
\label{fig:ep_nl_results_fixed}
\end{figure}

\begin{figure}[htb!]\centering
  \includegraphics[scale=0.6]{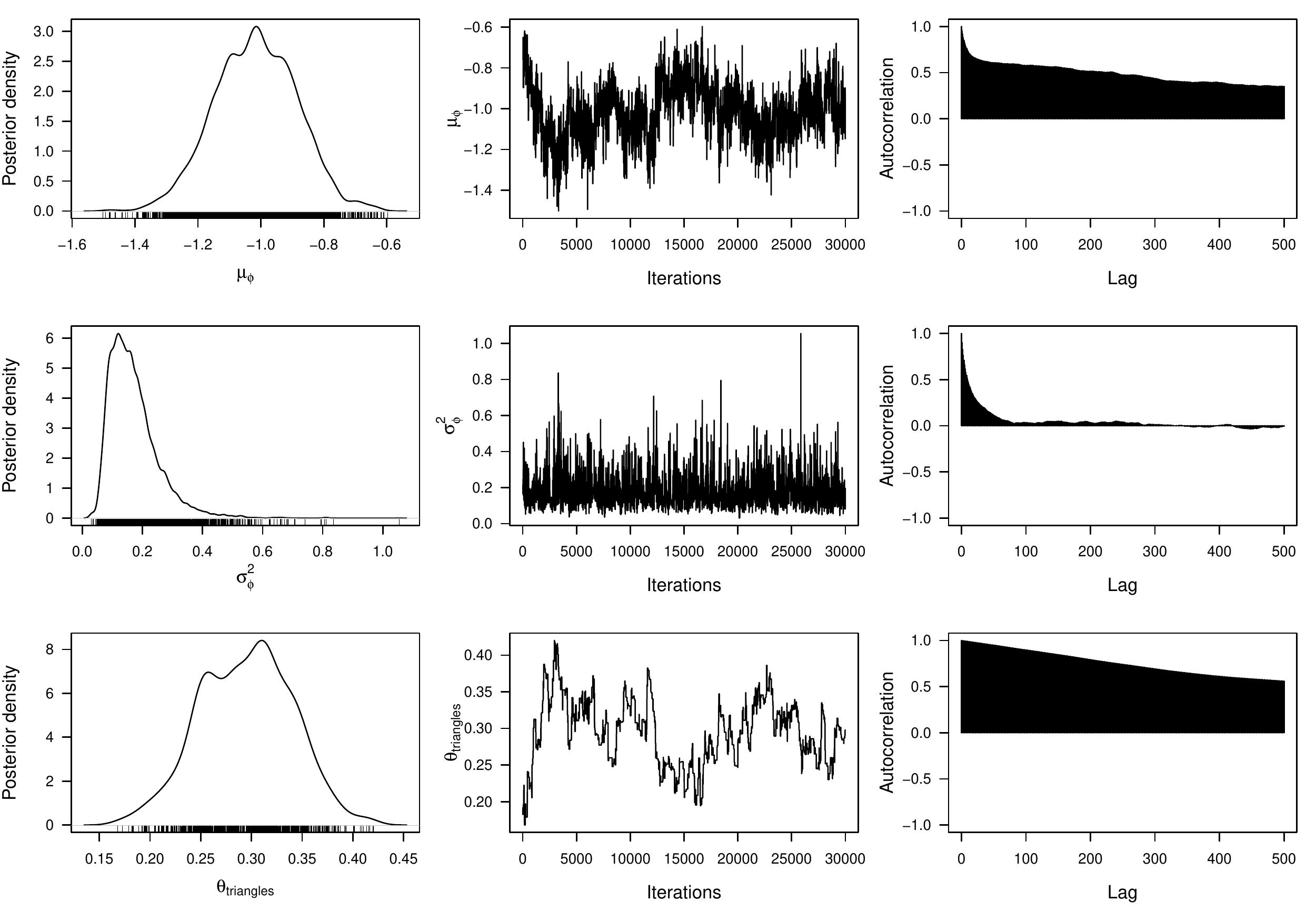} 
\caption{Posterior densities, trace plots, and autocorrelation for the mixed model with nodal random and triangular effects for the European parliament data.}
\label{fig:ep_nl_results_mixed}
\end{figure}

We fitted the same two nested models to the data as for the previous example: a standard ERGM with edges and triangles as sufficient statistics, and a model with nodal random effects and the triangle statistic. The number of iterations was also equivalent. \\
Figures \ref{fig:ep_nl_results_fixed} and \ref{fig:ep_nl_results_mixed} show the results, which are also summarised in Table \ref{tab:ep_nl_results}. For the mixed model we get a very low acceptance rate for the triangle effect and very high autocorrelations for the triangle effect and the mean parameter $\mu_\phi$. The later could possibly be solved by thinning out the chain. 

\begin{table}[htb!]\centering
\caption{\label{tab:ep_nl_results} Model fitting results for the European parliament  data.}
\begin{tabular}{llcccc}
\hline\\[-1ex]
Model type & Parameter & Post. mean & Post. Sd. & Acceptance rate & Note \\[1ex]
\hline\hline\\[-1ex]
\multirow{2}{*}{fixed} & $\theta_\text{edges}$ & -1.73 & 0.17 & \multirow{2}{*}{0.13}&\\
& $\theta_\text{triangles}$ & 0.26 & 0.04 \\[1ex]
\hline\\[-1ex]
\multirow{3}{*}{mixed} & $\mu_\phi$ & -1.02 & 0.13 & 0.11 &\\
& $\sigma^2_\phi$ & 0.15 & 0.08 & 0.15 & *\\
& $\theta_\text{triangles}$ & 0.29 & 0.05 & 0.02 & \\[1ex]
\hline\\[-1ex]
\multicolumn{6}{p{14cm}}{\footnotesize{* For $\sigma^2_\phi$ the posterior mean is calculated based on the logarithmized values and than transformed back to the scale of $\sigma^2_\phi$ (this leads to the geometric mean) due to the non-symmetric posterior density in this case.}}
\end{tabular}
\end{table}

Nevertheless, the focus in this example is on model selection. The computed log Bayes factor is -13.9 and clearly indicates that the model without nodal random effects is preferable in this situation. Apparently, here we have a network dataset where there is no benefit in including nodal random effects into the model. This corresponds to the rather small estimate for the variance of the nodal random effects $\sigma^2_\phi$. The resulting Bayes factor shows that it is not the case that the model with more parameters is always selected. This can also be seen from the simulation results in the following subsection.

\subsection{Simulation}
\label{sec:Simulation}

For the simulation study we used the following components based on two very simple, but different model generating processes, a nodal random effects only situation, i.e. the $p_2$ model, and structural effects only situation, i.e. the classical ERGM. For each setting we generated networks with 40 vertices, using again the simulation routines from the \texttt{ergm} package \citep{HunterHandcock:2008}. 
The first model (A) was the one with nodal random effects only, i.e.
\begin{align}\label{eq:sim-random}
\text{logit}\left[ \Prob\bigl(Y_{ij}=1 | Y_{kl}, (k,l) \neq (i,j); \vecphi\bigr) \right]
&= \phi_i + \phi_j, \\
& \text{with}\ \phi_i \sim \pnorm (\mu_\phi, \sigma^2_\phi),\quad \text{for } i = 1, \dots, n. \notag
\end{align} 
The parameter $\mu_\phi$ was constantly set to $\mu_\phi = -1$, so that the resulting network graphs tend to be rather sparse. For $\sigma^2_\phi$ we used values between $0$ and $1$. 
Model (B) was the standard ERGM with edges and 2-star statistics, and no nodal random effects, i.e. $\vectheta = (\theta_\text{edges}, \theta_\text{2-star})^t$ and
\begin{align}\label{eq:sim-ergm}
\text{logit}\left[ \Prob\bigl(Y_{ij}=1 | Y_{kl}, (k,l) \neq (i,j); \vectheta\bigr) \right]
&= \theta_\text{edges} + \theta_\text{2-star}\cdot \left[ \sum\limits_{k \neq j} y_{ik} + \sum\limits_{l \neq i} y_{jl} \right].
\end{align}
The parameter $\theta_\text{edges}$ was constantly set to $\theta_\text{edges} = -2$. This is equivalent to model (A) in the sense that $2 \cdot \mu_\phi = \theta_\text{edges}$, because $\theta_\text{edges}$ is a parameter on a per link basis, $\mu_\phi$ is on a per node basis and one needs two nodes to form a link. For $\theta_\text{2-star}$ we used values between $0$ and $0.05$. This value needs to be small, i.e. close to zero, because otherwise we only generate full or empty graphs if the value is negative, see also \cite{Schweinberger:2011}.\\
For each of the resulting parameter combinations in model (A) and model (B) we generated 50 networks. \\
For the chosen settings the resulting 40 node networks seem to be reasonable. We get an average network density between 0.11 and 0.30 for the different settings.\\
Note that setting $\sigma^2_\phi = 0$ in model (A) and $\theta_\text{2-star} = 0$ in model (B) leads to a simple Bernoulli network, which can be seen as a null model. \\
Similarly to the karate data example we fitted two nested models to each of the simulated networks: a standard ERGM with edges and 2-stars as sufficient statistics, and a model with nodal random effects and the 2-star statistic. Again this step was followed by computing a Bayes factor to compare the model with nodal random effects to the one with structural effects only. 

\begin{figure}[htb!]\centering
\includegraphics[scale=0.65]{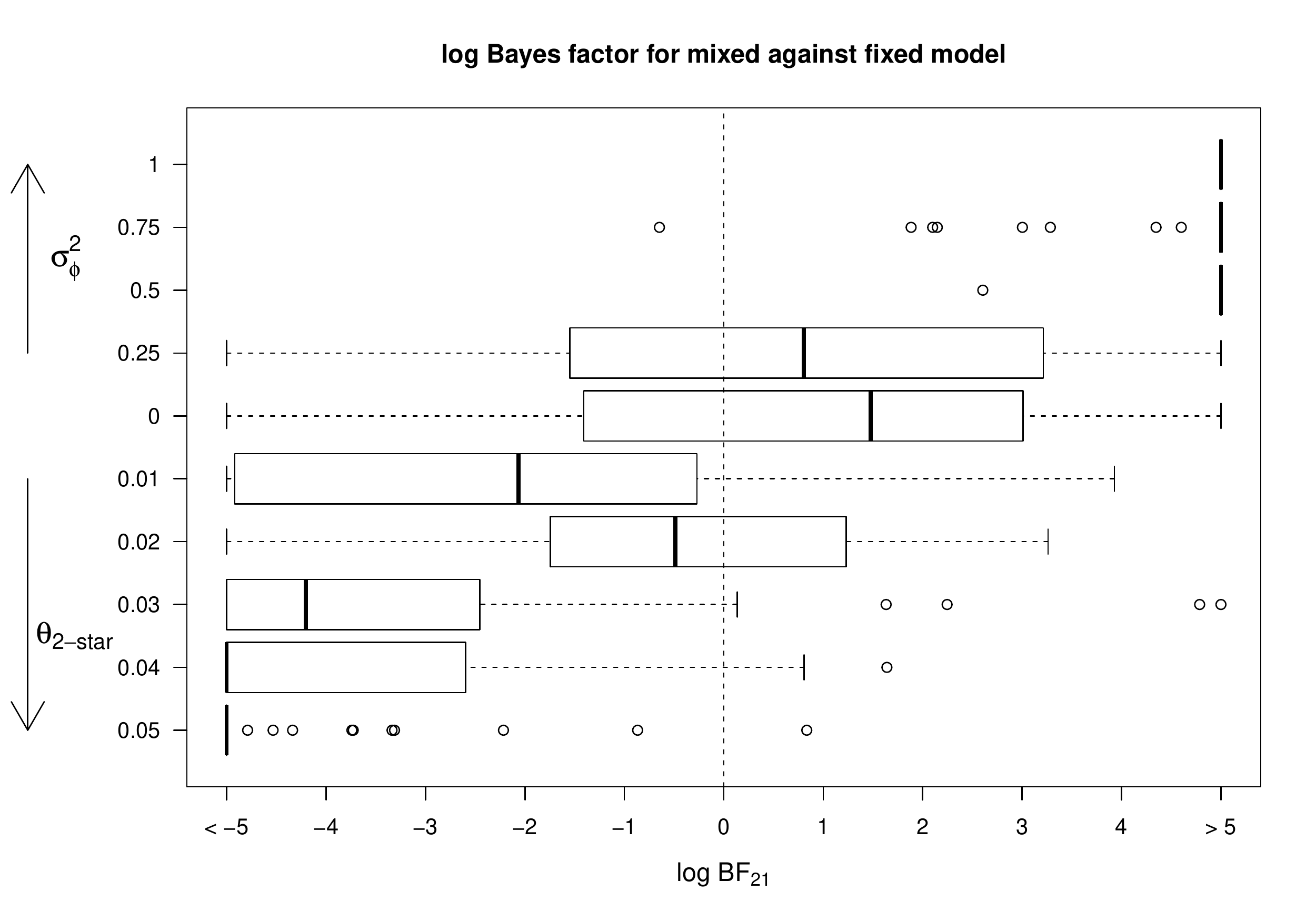}
\caption{\label{fig:sim_bf_all} Resulting log Bayes factors for the mixed model against the fixed model for different simulation settings. The annotation on the y-axis shows which was the underlying true model, a model with nodal random effects only in the direction of $\sigma^2_\phi$, and a model with edges and 2-stars in the direction of $\theta_\text{2-star}$.}
\end{figure} 

Figure \ref{fig:sim_bf_all} shows boxplots of the resulting log Bayes factors for the different settings. For the plot the log Bayes factors were cut at values of -5 and 5 because some were really small or really large. These cutting values were chosen following \cite{KassRaftery:1995}. More detailed information, especially on the range of the simulation results is given Table \ref{tab:sim_results}. For the null model of a pure Bernoulli network the log Bayes factor can point in either one of the directions, the same is more or less true for only small deviations from this null model. The general impression is, that the more extreme the underlying setting becomes the sooner the log Bayes factor points into the correct direction.\\
Most importantly the results of the simulation show that our model selection works with respect to the size of the competing models. It is not the case that the model with more parameters, which is the model with nodal random effects, is always preferred.

\begin{table}[htb!]\centering
\caption{Resulting Bays factors (mixed model against fixed model) for the simulation from setting (A) a nodal random effects only situation, and setting (B) a classical ERGM with edges and 2-star statistics. Each setting was run 50 times, except for the Bernoulli setting, which had $2 \cdot 50$ runs.}
\label{tab:sim_results}
\begin{tabular}{clcrrrrrr}
\hline \\[-1ex]
& & average & \multicolumn{6}{c}{log Bayes factor for mixed against fixed model} \\[1ex]
\multicolumn{2}{c}{Setting} & nw density & min & max & \% <-5 & \% < 0 & \% > 0 & \% > 5 \\[1ex]
\hline\hline \\[-1ex]
(A) & $\sigma^2_\phi = 1$ 		& 0.23 & 13.03 & 137.64 & 0 & 0 & 100 & 100 \\  
random & $\sigma^2_\phi = 0.75$ 	& 0.11 & -0.65 & 498.53 & 0 & 2 & 98 & 84 \\ 
effects & $\sigma^2_\phi = 0.5$ 	& 0.16 & 2.60 & 350.05 & 0 & 0 & 100 & 98 \\ 
& $\sigma^2_\phi = 0.25$ 		& 0.15 & -7.73 & 292.34 & 6 & 34 & 66 & 20 \\[1ex]
\hline\\[-1ex]
\multicolumn{2}{c}{Bernoulli network} & &&&&&&\\
\multicolumn{2}{c}{$\sigma^2_\phi = \theta_\text{2-star} = 0$} & 0.13 &-7.80 & 10.27 & 4 & 37 & 63 & 4 \\[1ex]
\hline\\[-1ex] 
& $\theta_\text{2-star} = 0.01$ 			& 0.13 & -14.56 & 3.93 & 24 & 76 & 24 & 0 \\ 
(B)& $\theta_\text{2-star} = 0.02$ 		& 0.14 & -144.23 & 3.26 & 10 & 54 & 46 & 0 \\ 
fixed & $\theta_\text{2-star} = 0.03$ 	& 0.16 & -25.24 & 51.34 & 44 & 88 & 12 & 2 \\ 
effects & $\theta_\text{2-star} = 0.04$ 	& 0.20 & -240.76 & 1.64 & 64 & 94 & 6 & 0 \\ 
& $\theta_\text{2-star} = 0.05$ 			& 0.30 & -218.07 & 0.83 & 80 & 98 & 2 & 0 \\ [1ex]
\hline
\multicolumn{9}{p{15cm}}{\footnotesize{Note: For setting (A) we set $\mu_\phi = -1$, and for setting (B) $\theta_\text{edges} = -2$, so that $\mu_\phi = 2 \cdot \theta_\text{edges}$.}}
\end{tabular}
\end{table}

\section{Discussion and Summary}
\label{sec:Discussion}

Statistical modelling of network data, with few exceptions, for example, \cite{KrivitskyHandcock:2009}, implicitly assumes that the local structure of the network is homogeneous. In particular, this implies that well studied phenomena, such as a small-world networks, \cite{milgram1967small}, \cite{watts1998collective}, where shortest path lengths between two nodes in the network tend to be very small and scale-free networks, where few nodes have unusually high degree, are not appropriately modelled using the standard statistical modelling approaches. This is particularly true for Exponential Random Graph Models. \\
Here our extension of the Exponential Random Graph Model (ERGM) avoids the assumption of nodal homogeneity. By adding nodal random effects to the model we get a flexible tool to model heterogeneity in the network which is not captured in available (nodal) covariates otherwise. Using the Bayesian framework for ERGMs proposed by \cite{CaimoFriel:2011} allows us to add this random effects extension to the model in an elegant and rather straightforward manner. 
Estimating Bayes factors enables us to handle the problem of model selection associated with this modelling task. The resulting estimates for the two data examples seem to be reasonable. \\
Furthermore, the small simulation study in the previous section suggests that in general the Bayes factor approach seems to work and even though a mixed model with nodal random effects has more parameters than its fixed equivalent it is not systematically preferred in the model selection.\\ 

We should note that the approach which we have introduced is computationally intensive. A promising avenue of research to address this issue is to explore approximations of the likelihood function using composite likelihoods, of which the pseudolikelihood approximation \cite{FrankStrauss:1986} is an antecedent. We refer the reader to \cite{varin-reid-firth11} for a recent review of composite likelihoods. We are currently engaged in work in this direction.

\subsubsection*{Acknowledgements}
We gratefully acknowledge the data provision by Paul W. Thurner.\\
 
The Insight Centre for Data Analytics is supported by Science Foundation Ireland under Grant Number SFI/12/RC/2289.
Nial Friel's research was also supported by an Science Foundation Ireland grant: 12/IP/1424.

\clearpage

\begin{appendix}
\section{Laplace approximation}\label{app:sec:laplace}

The likelihood in the mixed effects model marginalized over the random effects $\vecphi$ is
\begin{align}\label{eq:marlik_ext}
f( \by | \vectheta, \mu_\phi, \sigma^2_\phi )
&= 
\displaystyle\int \frac{ \exp\left\{ \vectheta^t s( \by ) + \vecphi^t t( \by ) \right\} }{ \kappa( \vectheta, \vecphi ) } \cdot p( \vecphi | \mu_\phi, \sigma^2_\phi ) \ \mbox{d} \vecphi \notag\\
&= 
\displaystyle\int \frac{ \exp\left\{ \vectheta^t s( \by ) + \vecphi^t t( \by ) \right\} }{ \kappa( \vectheta, \vecphi ) } \cdot \frac{ 1 }{ (2\pi)^\frac{n}{2} \left| \sigma^2_\phi I_n\right|^\frac{1}{2} } \cdot \notag\\
& \quad\quad 
\cdot \exp\left\{ - \frac{ 1 }{ 2\sigma^2_\phi } ( \vecphi - \mu_\phi \mathds{1}_n )^t  ( \vecphi - \mu_\phi \mathds{1}_n ) \right\} \mbox{d} \vecphi \notag\\
&= 
\frac{ \exp\left\{ \vectheta^t s( \by ) \right\} }{ (2\pi \sigma^2_\phi )^\frac{n}{2} } \cdot
\notag\\
& \quad\quad 
\cdot \displaystyle\int \exp\left\{ \vecphi^t t( \by ) - \frac{ 1 }{ 2\sigma^2_\phi } ( \vecphi - \mu_\phi \mathds{1}_n )^t  ( \vecphi - \mu_\phi \mathds{1}_n ) - \log( \kappa( \vectheta, \vecphi ) ) \right\}  \mbox{d} \vecphi.
\end{align}
The integral in equation (\ref{eq:marlik_ext}) is approximated around the point $\widehat{\vecphi}$ using a Laplace type approximation 
\begin{align}\label{eq:laplace_approx}
\displaystyle\int \exp\left\{ h( \vecphi ) \right\}  \mbox{d} \vecphi \ &\approx \ \exp\left\{ h( \widehat{\vecphi} ) \right\} (2\pi)^\frac{n}{2} \left| \Sigma \right|^{-\frac{1}{2}},
\end{align}
where 
\[
h( \vecphi ) = \exp\left\{ \vecphi^t t( \by ) - \frac{ 1 }{ 2\sigma^2_\phi } ( \vecphi - \mu_\phi \mathds{1}_n )^t  ( \vecphi - \mu_\phi \mathds{1}_n ) - \log( \kappa( \vectheta, \vecphi ) ) \right\} 
\]
and 
\begin{align*}
\Sigma 
&= 
\frac{ \partial^2 h( \widehat{\vecphi} ) }{ \partial \widehat{\vecphi} \partial \widehat{\vecphi}^t } \\
&= 
-\frac{1}{\sigma^2_\phi} I_n - \frac{ \partial^2 }{ \partial \widehat{\vecphi} \partial \widehat{\vecphi}^t } \log( \kappa( \widehat{\vectheta}, \vecphi ) ) \\
&= 
-\frac{1}{\sigma^2_\phi} I_n - \text{Cov}( t(\bY), t(\bY)^t | \widehat{\vecphi}, \vectheta ).
\end{align*}
The matrix $\text{Cov}( t(\bY), t(\bY)^t | \widehat{\vecphi}, \vectheta )$ denotes the covariance matrix of the vector of degree statistics $t(\bY)$ and can be estimated via simulated networks using the parameters $\widehat{\vecphi}$ and $\vectheta$. These networks are drawn in the same way as the auxiliary networks needed for the exchange algorithm described in Section \ref{sec:Bayesian}. \\
We assume that the posterior mode is close to the maximum likelihood estimator. The two are identical if the prior distributions are non-informative. This is not the case here, but we are assuming flat prior distributions and therefore the two should be reasonably close to each other. For reasons of simplicity, we use the posterior mean as value for $\widehat{\vecphi}$. \\
Combining equation (\ref{eq:marlik_ext}) with equation (\ref{eq:laplace_approx}) yields
\begin{align}
f( \by | \vectheta, \mu_\phi, \sigma^2_\phi )
&\approx 
\frac{ \exp\left\{ \vectheta^t s( \by ) \right\} }{ \kappa( \vectheta, \widehat{\vecphi} ) } \widehat{f}_\text{Laplace}( \by | \widehat{\vecphi}, \mu_\phi, \sigma^2_\phi ),
\end{align}
with
\[
\widehat{f}_\text{Laplace}( \by | \widehat{\vecphi}, \mu_\phi, \sigma^2_\phi ) 
= \sigma^{-n}_\phi \exp\left\{ \vecphi^t t( \by ) - \frac{ 1 }{ 2\sigma^2_\phi } ( \vecphi - \mu_\phi \mathds{1}_n )^t  ( \vecphi - \mu_\phi \mathds{1}_n ) \right\} \left| \Sigma \right|^{-\frac{1}{2}}.
\]

\end{appendix}

\clearpage

\bibliography{LaTeX/literatur}

\end{document}